\documentclass[prd,preprint,eqsecnum,amsfonts,amssymb,amsmath,floatfix,nofootinbib]{revtex4}

\usepackage{bm}
\usepackage{slashed}
\usepackage{graphicx}
\usepackage{MnSymbol}
\usepackage{braket}
\usepackage{appendix}
\usepackage{float}
\usepackage{mathtools}

\DeclareMathOperator{\sgn}{sgn}
\makeatletter
\newcommand{\raisemath}[1]{\mathpalette{\raisem@th{#1}}}
\newcommand{\raisem@th}[3]{\raisebox{#1}{$#2#3$}}
\makeatother

\begin{document}

\title{Experimental signatures of an alternative supersymmetry}

\author{Roland E. Allen}

\affiliation{Physics and Astronomy Department \\ Texas A\&M University, College Station, Texas 77843, USA}

\begin{abstract}
There are at least three physical arguments for some form of supersymmetry, based on experiment and observation, but conventional supersymmetry (SUSY) has not been observed up to surprisingly high experimental limits. Here we consider a radically different version, with initial bosonic fields in $32=16+\overline{16}$ (primitive sfermion) and $10=5+\overline{5}$ (primitive Higgs-related) representations of Spin(10) which do not satisfy Lorentz invariance. In the extremely early universe there is a reformation of these fields to achieve a stable Lorentz-invariant vacuum with two varieties of physical scalar-boson fields -- standard fields $\phi$ and fields $\varphi$ of a new kind. There are then two  possible scenarios: If sfermion fields are in the $\phi$ sector, the present description leads back to standard physics, including the standard model, SO(10) grand unification, and conventional SUSY. But if sfermion fields belong to the $\varphi$ sector, the predictions for production and decays of sparticles are dramatically different, potentially explaining their previous nonobservation. The masses of scalar bosons are still protected from enormous radiative corrections, gauge unification can be achieved, and there is a lowest-mass superpartner as a dark matter candidate — although it is presumed to be less abundant than the $\approx 70$ GeV candidate we introduced earlier in this same general context. Calculations by Shankar, Tallman, and Martinez in separate papers explore the possibilities for detection in future colliders, beginning with the high-luminosity LHC. 

\end{abstract}

\maketitle

\section{\label{introduction}Introduction}

There are at least three compelling \textit{physical} arguments for some form of supersymmetry~\cite{Baer-Tata,Dreiner,Nath,Mohapatra,Drees,Binetruy,Djouadi,Kane-susy,Haber-Kane,Arkani-Hamed-Giudice,Ellis-Olive,Baer-Barger-Tata,Roszkowski-2018,Baer-Barger-2020,Tata-2020,3,pdg} -- defined in the broadest sense~\cite{quote1} as a symmetry in which each fermion is matched by a boson with the same gauge quantum numbers, and each boson by a fermion, with their coupling constants also matched: The Higgs mass can be protected from radiative corrections that would lift it to absurdly large values; unification of gauge coupling constants can be achieved at high energy; and, assuming R-parity conservation, the lightest superpartner (LSP) will be a dark matter constituent. 

An additional argument is \textit{aesthetic} rather than physical: the mathematical beauty of conventional supersymmetry (SUSY), with a super-Poincar\'{e} algebra where
\begin{align}
\left\{ Q_{\alpha},Q^{\dag}_{\dot{\beta}} \right\} = 2  \sigma^{\mu} _{\alpha \dot{\beta}} P_{\mu} 
\label{Q}
\end{align}
in a standard notation (or $\left\{ Q_{\alpha},\overline{Q}_{\beta} \right\} = 2  \gamma^{\mu} _{\alpha \beta} P_{\mu} $ in the 4-component version), and the elegance of its extensions up to supergravity, string theory, and beyond.

However, Nature is not required to respect human aesthetic preferences, and we are now at a point where there is considerable skepticism about the the viability of any version of supersymmetry, following the exclusion at the LHC of expected superpartners in the most favored range of masses. Some skepticism was already being expressed more than 25 years ago, even by speakers at supersymmetry conferences (as witnessed by the present author at e.g. SUSY 97), but the lack of evidence for SUSY is now often perceived to represent an unfolding crisis in fundamental physics. 

One of the most recent attempts to resolve this crisis
%postulates that the dark matter consists entirely of axions~\cite{all-axion} -- a well-known but still hypothetical particle -- or 
employs a statistical analysis in the string theory multiverse with $10^{500}$, $10^{272,000}$, or more universes~\cite{landscape}. Other attempts involve searching for portions of narrow strips in the parameter space that are not yet ruled out by experiment~\cite{Ellis-Olive}; ignoring the problem of naturalness in the hierarchy problem~\cite{Arkani-Hamed-Giudice} (or again assuming that it can be resolved through anthropic arguments); or postulating that normal physics is truncated far below the Planck scale~\cite{large}.

In addition to the tension of SUSY with experiment, there are also theoretical impediments. For example, breaking conventional SUSY has been a central and unsolved problem for more than half a century. The difficulty ultimately results from (\ref{Q}), which implies that a mechanism for breaking conventional SUSY must \textit{increase} the energy~\cite{Fayet,Witten}, in contrast to the normal symmetry breakings elsewhere in physics which \textit{lower} the energy. This problem is ameliorated in the various versions of conventional supergravity, but these introduce further complications and difficulties, and there is still no completely successful or generally accepted solution.

In this paper we consider an alternative and radically unconventional form of supersymmetry  -- called susy here to avoid confusion -- which is compatible with the physical motivations listed above, but which redirects aesthetic preferences: There is a superpartner for each standard model particle, the Higgs mass is protected as usual from a nonlogarithmic divergence at high energy, unification of nongravitational coupling constants can be achieved, and the LSP is a dark matter constituent. This last feature, however, is now placed in a multi-component description, with a nonsupersymmetric WIMP like that of our previous papers~\cite{DM2021a,DM2021b,DM2022a,DM2022b,DM2025} assumed to be the dominant constituent; this related issue will also be addressed below. Many of the other aspects of conventional supersymmetry also remain in the present formulation, but some of the most prominent experimental signatures are dramatically changed in the version that is called scenario 2 below.

In the present description, the initial susy is broken by the requirement of a stable vacuum with Lorentz invariance, as primitive fields are transformed into physical fields for scalar bosons (including sfermions and Higgs-related particles). 

For concreteness, SO(10) grand unification~\cite{Nath,Mohapatra,SO(10)-1975,Nanopoulos-1984,Nanopoulos-1979,ross,Barger,Ozer,Pernow}  is assumed, with the actual group being Spin(10). 
As will be seen below, a full $32=16+\overline{16}$ spinorial representation is required for an extension of each generation of standard model fermions and their bosonic partners (sfermions), with a $10=5+\overline{5}$ vectorial  representation for each set of Higgs-related particles and their fermionic partners (higgsinos). (With all fields initially left-handed, those of the 16 and $\overline{16}$, or $5$ and $\overline{5}$, are independent.) For simplicity of nomenclature, even the color triplets of a 5 and $\overline{5}$ are called ``Higgs-related''.

The complete transformation from primitive to physical fields requires two sets of steps: In the appendix, the fields are first rearranged and rescaled, while their quantum numbers are left unchanged. In Section~\ref{scalars} the fields are then combined to achieve scalar boson fields consistent with a stable vacuum having Lorentz invariance. 

Physically this set of transformations is viewed as automatically occurring when, in the extremely early universe, the bosonic fields in an initially unstable vacuum reform and reorganize themselves -- becoming the physical fields revealed by these transformations -- in order to achieve the stability of the current vacuum. 

The joining of primitive fields to satisfy a required symmetry has various well-known precedents. For example, left-and right-handed Weyl fields must be joined to form a massive Dirac or Majorana field, in order to achieve Lorentz invariance, and two real fields must be joined to form the real and imaginary parts of an ordinary charged scalar field, to achieve gauge invariance.

The final fields $\phi$ and $\varphi$ are amplitude modes of penultimate 4-component fields $\Phi$ and $\overline{\Phi}$. These final fields describe physical excitations at accessible energies. corresponding to observable particles, and the 4-component fields are underlying fundamental fields in the vacuum. There are are many analogies in condensed matter physics, in which the low-energy excitations have a very different character than the original fundamental fields. The best-known example in high energy physics is the transformation of the fundamental $A^1_{\mu}$, $A^2_{\mu}$, $A^3_{\mu}$, and $B_{\mu}$ fields into the modified fields of a stable vacuum after Higgs condensation (the $W^+_{\mu}$, $W^-_{\mu}$, $Z^0_{\mu}$, and photon fields). In supersymmetry, higgsinos and electroweak gauginos must similarly be combined to from chargino and neutralino mass eigenstates.

After a general discussion of the initial primitive fields and their initial supersymmetry, in Section~\ref{general}, the $16 + \overline{16}$ and $5+\overline{5}$ representations are treated together in Section~\ref{scalars}.
The experimental consequences are discussed in Sections~\ref{physical} and \ref{conclusion}, with a brief mention of the calculations and experimental predictions reported in separate papers~\cite{DM2025,susy2025b}.

\section{\label{general}Primitive fields and their initial supersymmetry}

We begin with a full $10=5+\overline{5}$ vectorial or $32=16+\overline{16}$ spinorial representation of Spin(10). In the radically unconventional version of supersymmetry proposed here, each fermionic field $\psi^r _{f}$  is initially matched with a primitive (unphysical) bosonic field $\psi^r _{b}$:
\begin{align}
\psi^r =\left( 
\begin{array}{c}
\psi^r_b \\ 
\psi^r_f
\end{array}
\right) 
\quad \quad \mathrm{or} \quad \quad
\psi =\left( 
\begin{array}{c}
\psi _{b} \\ 
\psi _{f} 
\end{array}
\right) 
\label{primitive}
\end{align}
with the initial action
\begin{align}
S_{fb}&= \int d^{4}x \, i \psi ^{\dagger } \, \overline{\sigma} ^{\mu} D_{\mu } \, \psi  = S _f + S_b \label{eq700} \\
 S _f &= \int d^{4}x \, i \psi _{f}^{\dagger } \, \overline{\sigma} ^{\mu} D_{\mu } \, \psi _{f} 
\quad , \quad
S_b  = \int d^{4}x \, i \psi _{b}^{\dagger } \, \overline{\sigma}  ^{\mu} D_{\mu } \, \psi _{b}   \label{eq701} 
\end{align}
where $D_{\mu } = \partial_{\mu} - i A_{\mu}$, $A_{\mu}=A_{\mu}^i t^i$. 
The notation and conventions are further explained in the appendix. In particular, $\psi^r _{f}$ and $\psi^r _{b}$ are both 2-component Weyl spinors having the same gauge quantum numbers; 
 the SO(10) coupling constant is absorbed into $A_{\mu}$; the gauge generators $t^i$ are treated as operators; and, as evident in (\ref{eq700}) or  (\ref{eq701}), all the fields of $ \psi _{f} $ and $\psi _{b} $ are initially left-handed. 
 
 The Coleman-Mandula theorem~\cite{Coleman-Mandula} -- which implies that the symmetries described by ordinary Lie algebras (with only bosonic elements) cannot mix fields of different spins -- is trivially satisfied with the present modified version of supersymmetry, in which the initial fermionic and bosonic fields have the same form and action.  In conventional SUSY, this theorem is evaded via the extension to a graded algebra with both bosonic and fermionic elements. Here it is satisfied because the primitive susy transformations involving the primitive fields manifestly do not mix different spin states.
 
 The $\psi _{b}^r$ are clearly unphysical because they violate the spin-statistics connection required by Lorentz invariance, which we will assume to be required for a stable vacuum. In the extremely early universe, therefore, the vacuum must (through rapid dynamical processes) reconfigure itself to support transformed fields that are consistent with Lorentz invariance.
 
 In the appendix and the following section it is shown that this can be achieved in a natural way:  
 First, in the appendix, half the fields of $\psi _{f}$ and $\psi _{b}$ in each full representation are converted to right-handed fields, with a conventional result for fermions but a minus sign acquired for bosons (and a different conversion scheme).
The resulting action for the bosons is given by (\ref{eq702})-(\ref{eq704}).
Then it is shown that we can transform the fields further to obtain
\begin{align}
\hspace{-0.2cm}S_b &= S_{\phi} + S_F \\
S_{\phi}  &=\int d^{4}x \, \left( \phi_{\uparrow} ^{\dag } \, B \, \phi_{\uparrow}
+ \phi_{\downarrow} ^{\, \dag } \, B \, \phi_{\downarrow}  \right) 
=\int d^{4}x \, \sum_{r} \left( \phi^{r \, \dag} _{\uparrow}  B \, \phi^{r} _{\uparrow} 
+ \phi^{r \, \dag} _{\downarrow}  B \, \phi^r _{\downarrow}  \right) \label{B33x} \quad , \quad r=1,2, ..., N \\
 B  &= D^{\mu } D_{\mu } 
 \label{B}
 \end{align}
where N= 16 or 5, the notation is further defined in the appendix, and $S_F$ is the nondynamical action of the auxiliary fields.

This completes the first set of steps, in which the fields are rearranged and scaled. (More precisely, the Fourier components are rearranged and scaled, in order to form appropriate new fields with quantum numbers left unchanged.) It is clear, however, that  a second step is required to achieve physically acceptable fields, since excitations of $\phi_{\uparrow} ^{\, r}$ etc. would be spin 1/2 bosons, violating Lorentz invariance. In the next section, therefore, we combine the $\uparrow$ and $\downarrow$ fields to obtain proper scalar boson fields: the 4-component combined fields -- called $\Phi ^{r}$ and $\overline{\Phi }^{\, r}$ below -- and their amplitudes -- called $\phi ^{r}$ and $\varphi ^{\, r}$. Scalar bosons are interpreted as excitations of these amplitude modes, which are analogous to the Higgs/amplitude modes in superconductors~\cite{Varma1,Varma2,Shimano}.

\section{\label{scalars}Scalar boson fields, including redefined sfermions}

Proper scalar boson fields can be achieved in either of two ways:

\textbf{A conventional one-component (complex) scalar boson field} $\boldsymbol{\phi}$ can be obtained by combining two
fields $\phi_{\uparrow}^{\, r}$ and $\phi_{\downarrow}^{\, r}$ having the same gauge quantum numbers but opposite spins:
\begin{align}
\Phi ^{r}\left( x \right) &=\left( 
\begin{array}{c}
\phi_{\uparrow}^{\, r} \left( x \right) \\ 
 \phi_{\downarrow}^{\, r} \left( x \right)
\end{array} \right) 
\quad , \quad 
\phi_{\downarrow}^{\, r \, \dag} \left( x \right)  \phi_{\downarrow}^{\, r} \left( x \right)  = \phi_{\uparrow}^{\, r \, \dag} \left( x \right)  \phi_{\uparrow}^{\, r} \left( x \right)  \quad , \quad r=1,2, ..., N \
\end{align}
with
\begin{align}
S_{\phi}  = \int d^{4}x \, \sum_{r} \, \Phi ^{r \, \dag} \left( x \right)  B \; \Phi ^{r}\left( x \right) = \int d^{4}x \, \Phi ^{\dag} \left( x \right)  B \; \Phi \left( x \right)  \; .
\end{align}

We can define amplitude modes $\phi _i^{r}$ by
\begin{align}
\phi _i^{r}\left( x\right) &= \xi _{i}^{r\,\dag } \, \Phi^{r}\left( x\right)     \qquad  
\mathrm{with } \qquad   \xi _{i}^{r\,\dag }\,\xi _{i'}^r = \delta_{i i'} 
\label{e8x}
\end{align}
where $\xi ^r_{i}$ has $4$ constant components.  
Of the 4 orthonormal basis vectors $\xi^r_{i}$, we can choose $\xi_{3}^r$ and $\xi _{4}^r$  to be orthogonal to $\Phi ^{r}$, so that only $\phi _1^{r}\left( x\right)$ and $\phi _2^{r}\left( x\right)$ are nonzero.
For example, with the basis for spin up and down chosen such that, with all 4 components shown explicitly, 
\begin{align}
\Phi ^{r}\left( x \right) &=\left( 
\begin{array}{c}
\left[ \,\phi_{\uparrow}^{\, r} \left( x \right) \right]_1  \\ 
0 \\
0 \\
\left[ \, \phi_{\downarrow}^{\, r} \left( x \right) \right] _2
\end{array} \right) 
\end{align}
we can take the basis vectors to be 
\begin{align}
\xi ^r_{1} = \frac{1}{\sqrt{2}} \left( 
\begin{array}{c}
1 \\ 
0 \\
0 \\
1
\end{array} \right) \quad , \quad
\xi^r_{2} = \frac{1}{\sqrt{2}} \left( 
\begin{array}{c}
1 \\ 
0 \\
0 \\
- 1
\end{array} \right) \quad , \quad
\xi ^r_{3} = \frac{1}{\sqrt{2}} \left( 
\begin{array}{c}
0\\ 
1 \\
1 \\
0
\end{array} \right) \quad , \quad
\xi^r_{ 4} = \frac{1}{\sqrt{2}} \left( 
\begin{array}{c}
0\\ 
1 \\
-1 \\
0
\end{array} \right) \; .
\end{align}
There are then just the in-phase and out-of-phase amplitude modes:
\begin{align}
\Phi ^{r}\left( x\right) &=\phi _1^{r}\left( x\right) \,\xi ^r_{1} + \phi _2^{r}\left( x\right) \,\xi^r_{2} 
\end{align}
so that
\begin{align}
S_{\phi} &=\int d^{4}x \, \sum_{r} \, \left[ \phi _1 ^{r \, *} \left( x \right)  B \; \phi _1^{r} \left( x \right) 
+ \phi _2^{r \, *} \left( x \right)  B \; \phi _2^{r }\left( x \right)  \right] \\
&=\int d^{4}x \, \sum_{r} \, \left[ \phi _1^{r \, *} \left( x \right)  B \; \phi _1^{r} \left( x \right) 
+ \phi _2^{r \, c \, *} \left( x \right)  B \; \phi _2^{r \, c }\left( x \right)  \right] \\
&=\int d^{4}x \, \left[ \phi _1^{ \dag} \left( x \right)  B \; \phi _1 \left( x \right) 
+ \phi _2^{c\, \dag} \left( x \right)  B \; \phi _2^{ c}\left( x \right)  \right] \label{ampl}
\end{align}
where 
\begin{align}
\phi_i^{\, r \, c}  \left( x \right) = C \phi _i ^{\, r \, *} \left( x \right) \; . 
\end{align}
We have used the fact that
\begin{align}
\int d^{4}x \, \phi _i^{r \, *} \left( x \right)  B \; \phi _i^{r }\left( x \right)  =  \int d^{4}x \, \phi _i^{r \, c \, *} \left( x \right)  B \; \phi _i^{r \, c}\left( x \right) 
\end{align}
follows from a simpler version of the argument in (\ref{conj})-(\ref{eqA20}).
(When the conjugate fields are placed in an array, they are, of course, reordered so that each has its appropriate place in the gauge multiplet.)

For a $5 + \overline{5}$ representation, $\phi_1$ and $\phi _2^{c}$ each consist of 5 one-component scalar boson fields; and for a $16 + \overline{16}$ representation,  $\phi_1$ and $\phi _2^{c}$  each consist of 16 one-component scalar boson fields.

From a $5 + \overline{5}$ we obtain the usual two Higgs doublets of supersymmetry. 

For each of the three $16 + \overline{16}$ families we can obtain 16 sfermions to match the 16 fermions of the standard model, and we can also obtain 16 conjugate sfermions to match the additional 16 conjugate fermions of the $\overline{16}$ that are predicted in the present description. (Again, with all fermion fields initially left-handed, those of the 16 and $\overline{16}$ are independent.)
This possibility will be called scenario 1. The sfermions will then be conventional scalar bosons, and the experimental predictions will be essentially the same as those of conventional SUSY if we make the same assumptions regarding the couplings of scalar bosons to fermions etc. -- for example, with sfermion-Higgs couplings 
\begin{align}
\overline{\mathcal{L}}^{H}_{\phi} = - \, y_{r} ^2 \; \phi_H^{r \, *} \phi_H^{r} \; \phi ^{r \, *}  \phi ^{r}   \label{eq50}
\end{align}
where $y_{r} $ is the same as the Yukawa coupling for the corresponding fermion.

Conventional SUSY has so far proved unsuccessful, however, so in the remainder of the paper we will consider only the alternative scenario 2 described immediately below.

The prediction of three additional families of fermions and sfermions, from the $\overline{16}$ representations, is consistent with experiment and observation if the masses are large.

\textbf{Unconventional scalar boson fields} $\boldsymbol{\varphi}$ can be constructed by combining 
$\phi_{\uparrow}^{\, r}$ and a charge-conjugate field 
\begin{align}
\phi_{\downarrow}^{\, r \, c}  \left( x \right) = C \phi_{\downarrow}^{\, r \, *} \left( x \right)
\end{align}
having both opposite spin and opposite gauge quantum numbers:
\begin{align}
\overline{\Phi} ^{r} \left( x \right) &=\left( 
\begin{array}{c}
\phi_{\uparrow}^{\, r} \left( x \right) \\ 
 \phi_{\downarrow}^{\, r \, c} \left( x \right)
\end{array} \right) 
\quad , \quad 
\phi_{\downarrow}^{\, r \, c \, \dag} \left( x \right)  \phi_{\downarrow}^{\, r \, c} \left( x \right)  = \phi_{\uparrow}^{\, r \, \dag} \left( x \right)  \phi_{\uparrow}^{\, r} \left( x \right)  \label{same}\; .
\end{align}
This scenario 2 can be initially formulated for a full Spin(10) representation, but below we will consider the simplest final (low-energy) version, in which the relevant representations (after symmetry breakings) are the fundamental representations of the SU(3), SU(2), and U(1) subgroups of the standard model, using
\begin{align}
\{ t^j,t^{j'} \}=\frac{1}{N} \delta^{jj'} + d_{jj'j''} t^{j''}
\label{anticomm}
\end{align}
which holds for a fundamental representation of any $SU(N)$, where the $d_{jj'j''}$ are structure constants. 

To avoid cumbersome notation, from the above paragraph through the end of this section the $t^{j}$ are generators for an arbitrary SU(N), with
(\ref{anticomm}) giving
\begin{align}
 A^{\mu }  A_{\mu} =  A^{\mu \, j} t^j A_{\mu}^{j' } t^{j'} = \frac{1}{2} g^{\mu \nu}  A_{\mu }^{j } A_{\nu}^{j' } \{ t^j , t^{j'} \}  =  A^{\mu \, j} A_{\mu}^{j' }\left( \frac{1}{2N} \delta^{jj'} +  \frac{1}{2} d_{jj'j''} t^{j''} \right) \; .
\end{align}
The results below also hold for a $U(1)$ representation with $1/(2N) \rightarrow 1$.

We will also need, for the Fourier coefficients,
\begin{align}
 \phi_{\downarrow} ^{\, r \, c \, \dag } \left( p \right) \, t^j \; \phi_{\downarrow} ^{\, r \, c} \left( p \right) 
= - \; \phi_{\downarrow} ^{\, r \, \dag } \left( p \right) \, t^j \; \phi_{\downarrow} ^{\, r } \left( p \right) 
= - \; \phi_{\uparrow} ^{\, r \, \dag } \left( p \right) \, t^j \;  \phi_{\uparrow} ^{\, r} \left( p \right) 
\end{align}
which follows from (\ref{same}) because $\phi_{\downarrow} ^{\, r}  \left( p \right)$ (as defined here) and $\phi_{\uparrow}^{\, r} \left( p \right) $ have the same amplitude and gauge quantum numbers..

Since  
\begin{align}
\int d^{4}x \, \phi_{\downarrow} ^{\, r \, \dag }  \left( x \right)  B^{r} \phi_{\downarrow} ^{\, r}  \left( x \right)  = 
\int d^{4}x \, \phi_{\downarrow} ^{\, r \, c \, \dag }  \left( x \right)  B^{r} \phi_{\downarrow} ^{\, r \, c}  \left( x \right)
\end{align}
again follows from a simpler version of the argument in (\ref{conj})-(\ref{eqA20}), we have
\begin{align}
S_{\varphi}  &= \int d^{4}x \, \sum_{r} \left( \phi_{\uparrow} ^{\, r \,\dag } \left( x \right) \, B^{r} \, \phi_{\uparrow} ^{\, r} \left( x \right) + \phi_{\downarrow} ^{\, r \, c \, \dag } \, B^{r} \phi_{\downarrow} ^{\, r \, c} \left( x \right)  \right)  \label{invar}\\
%&= \int d^{4}x \, \sum_{r} \bigg[  \phi_{\uparrow} ^{\, r \,\dag } \left( x \right) \, \left(  \left(  \partial^{\mu}  - i A^{\mu \, j} t^j \right)\left(  \partial_{\mu}  - i A_{\mu}^{j' }t^{j'} \right)  - \bar{y}_{r} ^2 \, \phi_H^{r \, *} \phi_H^{r}  \right) \, \phi_{\uparrow} ^{\, r} \left( x \right) \nonumber \\
%& \hspace{2.2cm} + \phi_{\downarrow} ^{\, r \, c \, \dag } \left( x \right) \, \left(  \left(  \partial^{\mu}  - i A^{\mu \, j} t^j \right)\left(  \partial_{\mu}  - i A_{\mu}^{j' }t^{j'} \right)  - \bar{y}_{r} ^2 \, \phi_H^{r \, *} \phi_H^{r}  \right) \phi_{\downarrow} ^{\, r \, c} \left( x \right)    \bigg] \\
&= \int d^{4}x \, \sum_{r}  \bigg[ \sum_{p} \phi_{\uparrow} ^{\, r \,\dag } \left( p \right) e^{i p \cdot x } \, \left(  \partial^{\mu}  - i A^{\mu \, j} t^j \right)\left(  \partial_{\mu}  - i A_{\mu}^{j' }t^{j'} \right)   \, \sum_{p'} \phi_{\uparrow} ^{\, r} \left( p' \right) e^{i p' \cdot x } \nonumber \\
& \hspace{2.2cm} + \sum_{p}  \phi_{\downarrow} ^{\, r \, c \, \dag } \left( p \right) e^{i p \cdot x }  \, \left(  \partial^{\mu}  - i A^{\mu \, j} t^j \right)\left(  \partial_{\mu}  - i A_{\mu}^{j' }t^{j'} \right)   \sum_{p'} \phi_{\downarrow} ^{\, r \, c} \left( p' \right) e^{i p' \cdot x }    \bigg]   \\
&= {\cal V} \sum_{r} \sum_{p } \bigg[ \phi_{\uparrow} ^{\, r \,\dag } \left( p \right)  \, \left(  \left( i p^{\mu}  - i A^{\mu \, j} t^j \right)\left( i p_{\mu}  - i A_{\mu}^{j' }t^{j'} \right) - i \left( \partial^{\mu} A_{\mu}^{j' } \right) t^{j'}   \right) \, \phi_{\uparrow} ^{\, r}  \left( p \right)  \nonumber \\
&  \hspace{1.2cm} + \phi_{\downarrow} ^{\, r \, c \, \dag } \left( p \right)   \, \left(  \left( i p^{\mu}  - i A^{\mu \, j} t^j \right)\left( i p_{\mu}  - i A_{\mu}^{j' }t^{j'} \right)  - i \left( \partial^{\mu} A_{\mu}^{j' } \right) t^{j'}    \right) \phi_{\downarrow} ^{\, r \, c} \left( p \right)  \bigg]  \label{v1}\\
&={\cal V}  \sum_{r} \sum_{p} \bigg[ \phi_{\uparrow} ^{\, r \,\dag } \left( p \right)  \, \left( - p^{\mu} p_{\mu}  - A^{\mu j} A^j_{\mu}/\left( 2 \, N \right) \right)    \, \phi_{\uparrow} ^{\, r}  \left( p \right)  \nonumber \\
&   \hspace{2.2cm}+ \phi_{\downarrow} ^{\, r \, c \, \dag } \left( p \right)   \,  \left( - p^{\mu} p_{\mu}  - A^{\mu j} A^j_{\mu}/\left( 2 \, N \right)  \right)    \phi_{\downarrow} ^{\, r \, c} \left( p \right)    \bigg]  \label{v2}\\
&= \int d^{4}x \, \sum_{r} \phi_{\uparrow} ^{\, r \,\dag } \left( x \right) \, \left(   \partial^{\mu} \partial_{\mu}  - A^{\mu j} A_{\mu j}/\left( 2 \, N \right) \right)   \, \phi_{\uparrow} ^{\, r}  \left( x \right)   \nonumber \\
&   \hspace{2.2cm}+ \phi_{\downarrow} ^{\, r \, c \, \dag } \left( x\right)   \, \left(  \partial^{\mu} \partial_{\mu}  - A^{\mu j} A_{\mu j}/\left( 2 \, N \right)   \right) \phi_{\downarrow} ^{\, r \, c} \left( x\right)    \bigg]  \\
&= \int d^{4}x \, \sum_{r} \overline{\Phi}  ^{\, r \,\dag } {\cal B }^{r}\, \overline{\Phi}  ^{\, r}  \left( x \right)  \quad , \quad  {\cal B }^{r} = \left( \partial^{\mu}  \partial_{\mu}  - A^{\mu j} A^j_{\mu}/\left( 2 \, N \right) \right)  \label{calB} \\
&= \int d^{4}x \, \overline{\Phi} ^{\,\dag } \left( x \right) \, {\cal B}  \, \overline{\Phi}  \left( x \right)   
\end{align}
where the matrix elements of ${\cal B }$ are $\delta_{r,r'}{\cal B }^{r}$, ${\cal V}$ is a 4-dimensional normalization volume, and we have renamed $S_{\phi} \rightarrow S_{\varphi} $ in the present context.

We can again define amplitude modes $\varphi _i^{r}$ and $\varphi _2^{r}$, by
\begin{align}
\varphi _i^{r}\left( x\right) &= \zeta _{i}^{r\,\dag } \, \overline{\Phi} ^{r}\left( x\right)     \quad , \quad
\mathrm{with } \quad  \quad \zeta _{i}^{r\,\dag }\,\zeta _{i'}^r = \delta_{i i'} 
\label{e8y}
\end{align}
where $\zeta ^r_{i}$ has $4$ constant components.  
Of the 4 orthonormal basis vectors $\zeta^r_{i}$, we choose $\zeta _{3}^r$ and $\zeta _{4}^r$  to be orthogonal to $\overline{\Phi} ^{r}$, so that only $\varphi _1^{r}\left( x\right)$ and $\varphi _2^{r}\left( x\right)$ are nonzero.

Then (\ref{calB}) gives
\begin{align}
S_{\varphi}  &= \int d^{4}x \, \sum_{r} \overline{\Phi} ^{\, r \,\dag } {\cal B }^{r}\, \sum_i \zeta ^r_{i} \zeta ^{i \, \dag}_{r} \; \overline{\Phi} ^{\, r}  \left( x \right) \\
&= \int d^{4}x \, \sum_{r} \left[ \varphi_1 ^{\, r \, * }\left( x\right)  {\cal B }^{r}\, \varphi_1 ^{r}\left( x\right)   + \varphi_2 ^{\, r \, * }\left( x\right)  {\cal B }^{r}\, \varphi_2 ^{r}\left( x\right)  \ \right] \\
&= \int d^{4}x \, \sum_{r} \left[ \varphi_1 ^{\, r \, * }\left( x\right)  {\cal B }^{r}\, \varphi_1 ^{r}\left( x\right)   + \varphi_2 ^{\, r \, c \, * }\left( x\right)  {\cal B }^{r}\, \varphi_2 ^{r \, c}\left( x\right)   \right] \\
&= \int d^{4}x \, \sum_{r} \left[ \varphi_1 ^{\, \dag }\left( x\right)  {\cal B }^{r}\, \varphi_1 \left( x\right)   + \varphi_2 ^{\, c \, \dag }\left( x\right)  {\cal B }^{r}\, \varphi_2 ^{c}\left( x\right)  \ \right] \label{var}
\end{align}
since 
$\varphi_2 ^{r \, *} \left( x \right)  {\cal B }^{r}  \, \varphi_2 ^{r }\left( x \right) $ = $ \varphi _2^{r \, c \, *} \left( x \right)  {\cal B }^{r}  \, \varphi _2^{r \, c}\left( x \right) $ as before, with $\varphi_2 ^{r \, c}\left( x \right)  = C \varphi _2^{r \, *} \left( x \right)$.  For a $5 + \overline{5}$ representation, $\varphi_1$ or $\varphi _2^{c}$ each consist of 5 complex scalar boson fields; and for a $16 + \overline{16}$ representation,  $\varphi_1$ and $\varphi _2^{c}$ each consist of 16 complex scalar boson fields.

%There is, however, a qualification for the preceding statements: In general it is possible for a representation to have a mixture of $\phi $ and $\varphi$ mass eigenstates, since some of the $\phi_{\uparrow}^{\, r}$ and $\phi_{\downarrow}^{\, r}$ may be combined as in scenario 1 and the others as in scenario 2. 

The lowest mass $\varphi$ particle from a $5+\overline{5}$ representation is the dark matter candidate of our previous papers~\cite{DM2021a,DM2021b,DM2022a,DM2022b,DM2025} (although the derivation here differs from that in~\cite{DM2021a}). In earlier papers we have used the generic term \textit{higgsons} for the particles in the $\varphi$ sector that correspond to the Higgs-related particles in the $\phi$ sector, because they have the same coupling constants. In the simplest picture, there are two $5+\overline{5}$ multiplets, with one containing the same two Higgs doublets as conventional SUSY, and the other containing two corresponding higgson doublets. The lowest-mass neutral particle from the conventional $\phi$ pair of Higgs doublets is then the observed 125 GeV Higgs boson, and the lowest-mass neutral particle from the $\varphi$ pair is our predicted $\approx 70$ GeV dark matter WIMP.

From three $16 + \overline{16}$ families we again can obtain 16 sfermions to match the 16 fermions of the standard model, and  16 conjugate sfermions to match an additional 16 conjugate fermions of the $\overline{16}$. This possibility, called scenario 2, will be emphasized here. 

Written more explicitly, (\ref{var}) is
\begin{align}
S_{\varphi} &=\int d^{4}x \, \sum_{r} \bigg[ \varphi _1^{r \,  \dag} \left( x \right) \left[ \left(  \partial^{\mu} \partial_{\mu}  - A^{\mu j} A^j_{\mu}/\left( 2 \, N \right) \right)  \right]\, \varphi _1^r \left( x \right) \nonumber \\
& \hspace{2.3cm}+ \varphi _2 ^{r \, c\, \dag} \left( x \right) \left[  \left(  \partial^{\mu} \partial_{\mu}  - A^{\mu j} A^j_{\mu}/\left( 2 \, N \right) \right)  \right]\, \varphi _2^{r \, c}\left( x \right) \bigg] 
  \label{var2}
\end{align}
with, again, $N =$ dimension of the representation. 

The action for the fundamental fields is gauge-invariant in the usual way, as can be seen from (\ref{B}) and (\ref{invar}) in the form
\begin{align}
S_{\varphi} &= \int d^{4}x \, \sum_{r} \overline{\Phi}^{\, r \,\dag } \left( x \right)  D^{\mu } D_{\mu } \, \overline{\Phi}^{\, r} \left( x \right)   \label{invarx} \; .
\end{align}
Then it is immediately obvious that (\ref{var2}) is also gauge invariant: When $A^j_{\mu} \rightarrow A^{j \, \prime}_{\mu}$ and $\overline{\Phi}^{\, r}  \left( x\right) \rightarrow \overline{\Phi}^{r \, \prime}\left( x\right)$ in (\ref{invarx}),
\begin{align}
\varphi^{r}_{1}\left( x\right) \rightarrow \varphi^{r \, \prime}_{1}\left( x\right) = \zeta _{1}^{r \, \dag } \, \overline{\Phi}^{\, r \, \prime}\left( x\right)  \quad , \quad \varphi^{r}_{2}\left( x\right) \rightarrow \varphi^{r \, \prime}_{2}\left( x\right) = \zeta _{2}^{r \, \dag } \, \overline{\Phi} ^{\, r \, \prime}\left( x\right)  \; .
\end{align}

\section{\label{physical}Experimental consequences}

Scenario 1, defined above, yields essentially the same physical predictions as conventional SUSY, so all the statements and results in the remainder of this paper are entirely within the context of scenario 2.

For the SU(3)$\times$SU(2)$\times$U(1) fields of the standard model, if coupling constants are displayed rather than absorbed into the gauge potentials, (\ref{var2}) implies that the Lagrangian for the interaction of sfermions with gauge fields is
\begin{align}
\overline{\mathcal{L}}^{int} &=  - \varphi^{r\, \dag } \left( \, \sum_n \overline{g}_n^2 \, A_n^{\mu  j} A_{n \mu}^j  \, \right) \varphi^r \quad , \quad \overline{g}_{3}^2 = g_{3}^2 /6 , \quad \overline{g}_{2}^2 = g_2^2/4, \quad \overline{g}_{1}^2 = g_1^2
\label{eq30}
\end{align}
where $g_{3}$, $g_{2}$, $g_{1}$ are the original SU(3), SU(2), U(1) coupling constants. 
This form for the Lagrangian, and the developments below, hold when $\varphi^r$ is replaced by any of the above $\varphi_1^r$ or $\varphi _2^{r \, c }$.

When the original $SU(2) \times U(1)$ fields are rotated into those of the electroweak theory after symmetry breaking
\begin{align}
W^{\pm}_{\mu} = \frac{1}{\sqrt{2}} \left( A^1_{2 \mu } \mp i A^2_{2 \mu } \right) \;, \;
Z_{\mu} = \frac{1}{\sqrt{g_1^2 + g_2^2}} \left( g_2 A^3_{2 \mu } - g_1 A_{1 \mu } \right) \; , \;
\bar{A}_{\mu} = \frac{1}{\sqrt{g_1^2 + g_2^2}} \left( g_1 A^3_{2 \mu } + g_2 A_{1 \mu } \right) \nonumber
\end{align}
with the covariant derivative~\cite{peskin}
\begin{align}
D_{\mu }=\partial _{\mu } -i\frac{g}{\sqrt{2}}\left( W_{\mu}^{+}\tau^{+}+W_{\mu }^{-}\tau^{-}\right)   - i \frac{g}{\cos \theta _{w}}Z_{\mu }\left( \tau^{3}-\sin ^{2}\theta_{w}\,Q\right) -ieA_{\mu }\,Q  \label{e5}
\end{align}
where $\tau^{\pm} = \tau^1 \pm i \tau^2$, (\ref{e5}) and (\ref{eq30}) give 
\begin{align}
- \frac{g_s^2}{6} \,  {\cal A}^{\mu  i}  {\cal A} _{\mu}^i  \quad , \quad  - \frac{g^2}{2} \, W^{+ \mu } W^- _{\mu}  \quad , \quad - \frac{g_Z^2}{4}  \, Z^{\mu } Z_{\mu} \quad , \quad - \left( Qe \right)^2 \, \bar{A}^{\mu } \bar{A}_{\mu}   \label{eq43} 
\end{align}
for the strong, weak, and electromagnetic interactions respectively.
Here $g_s=g_3$ and $g=g_2$ are the usual strong and weak coupling constants, $g_Z=g/\cos \theta_W$, $\cos \theta_W = g_2/\sqrt{g_1^2 + g_2^2}$, $Q e$ is the electric charge, $\bar{A}_{\mu }$ is the electromagnetic vector potential, and ${\cal A}_{\mu } = {\cal A}_{\mu}^i T^i $ is the QCD gauge field containing gluon fields ${\cal A}_{\mu }^i$. 

The $\varphi^r$ (as mass eigenstates) are the final redefined sfermion fields of the present theory. 
We also have the  sfermion-Higgs couplings of (\ref{eq50}) (which are left unchanged by charge conjugation):
\begin{align}
\overline{\mathcal{L}}^{H}_S = - \,  y_{r} ^2 \; \phi_H^{r \, *} \phi_H^{r} \; \varphi ^{r \, *}  \varphi ^{r} \; . \label{eq50x}
\end{align}
It is obvious for which fields the various interactions of (\ref{eq43}) and (\ref{eq50x}) are relevant; for example, a left-handed squark experiences all of them.
%with a Lagrangian 
%\begin{align}
%\overline{\mathcal{L}}_{\widetilde{q}_L} = \varphi_{\widetilde{q}_L}^{\dag } \left( \partial^{\mu} \partial_{\mu}  - \frac{g_s^2}{6} \,  {\cal A}^{\mu  i}  {\cal A} _{\mu}^i   - \frac{g^2}{2} \, W^{+ \mu } W^- _{\mu}   - \frac{g_Z^2}{4}  \, Z^{\mu } Z_{\mu}  - \left( Qe \right)^2 \, \bar{A}^{\mu } \bar{A}_{\mu}  - y^2_{\widetilde{q}} \, \phi_H^0 \phi_H^0 \right) \varphi_{\widetilde{q}_L}  \label{98}
%|end{align}
%and a (left-handed) sneutrino experiences only the weak and Higgs interactions, with the Lagrangian
%\begin{align}
%\overline{\mathcal{L}}_{\widetilde{\nu}} = \varphi_{\widetilde{\nu}}^{\dag } \left( \partial^{\mu} \partial_{\mu}  - \frac{g^2}{2} \, W^{+ \mu } W^- _{\mu}   - \frac{g_Z^2}{4}  \, Z^{\mu } Z_{\mu}  - y^2_{\widetilde{\nu}} \, \phi_H^0  \phi_H^0 \right) \varphi_{\widetilde{\nu}}  \; .
%\label{99}
%\end{align}
Recall that the full Lagrangians have the gauge invariance demonstrated at the end of Section~\ref{scalars}.

\begin{figure}[H]
\begin{center}
\resizebox{0.5\columnwidth}{!}{\includegraphics{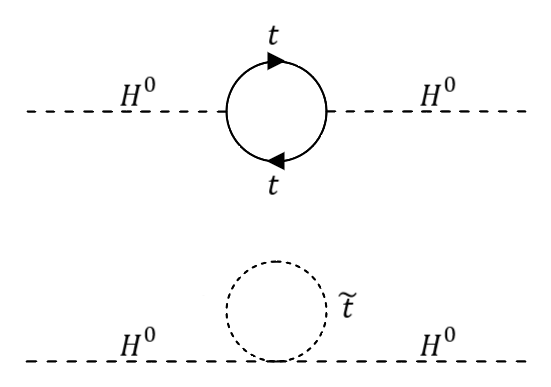}}
\resizebox{0.25\columnwidth}{!}{\includegraphics{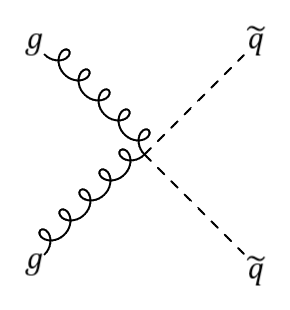}}
\end{center}
\caption{Left panel: Representative diagrams for contributions of fermion and sfermions -- in this case top quark and top squarks -- to quantum corrections of Higgs mass-squared. According to (\ref{eq50x}), the unconventional sfermions defined here will still provide the standard supersymmetric cancellation of  quadratric divergences, provided that all sfermions have masses not far above a few TeV. Examination of the relevant diagrams indicates that this cancellation holds for all processes in higher-order diagrams, since fermions and sfermions are coupled to both Higgs bosons and gauge bosons. 
%There will still be logarithmic corrections from fermions and sfermions, just as in conventional SUSY, but this is consistent with experiment for particles with masses again not far above a few TeV. 
Right panel: Many sfermion production processes are still allowed in the present scenario -- for example, production of squarks by direct gluon fusion, shown here.}
\label{squarks}
\end{figure}

As noted above, and indicated in Fig.~\ref{squarks}, the interactions of  (\ref{eq50x}) are still what is needed to give the usual supersymmetric cancellation of the quadratically-divergent radiative correction from fermion loops to the squared mass of the observed Higgs boson.
% reducing it to a manageable logarithmic divergence. 
Many of the conventional processes for squark and gluino production do not exist in the present scenario, because they involve first-order interactions. Many others, however, are still allowed -- for example, production of squarks by direct gluon fusion and higher-order vector-boson fusion, with the first of these depicted in Fig.~\ref{squarks}. 

A more dramatic difference is the result that the decays of squarks and gluinos -- which are of central importance in conventional searches for supersymmetry -- do not occur at all in the present scenario 2, because of the form of the interactions (\ref{eq43}) and (\ref{eq50x}): A single incident sfermion must always emerge after any process. These redefined sfermions (or their hadronized complexes) can then be minor components of the dark matter if they are stable: They are electrically neutral and colorless (with zero expectation value for every charge operator), as required for dark matter. But in the early universe, with typically stronger interactions and larger masses, most will annihilate more rapidly or else have exponentially lower thermal abundances than an approximately 70 GeV WIMP~\cite{DM2021a,DM2021b,DM2022a,DM2022b,DM2025} , so that their relic abundances should be substantially lower as they are thinned out in an expanding universe.
\begin{figure}[H]
\begin{center}
\resizebox{0.35\columnwidth}{!}{\includegraphics{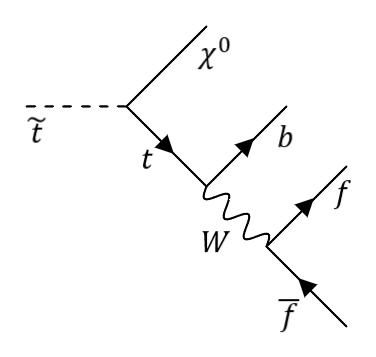}}
\resizebox{0.63\columnwidth}{!}{\includegraphics{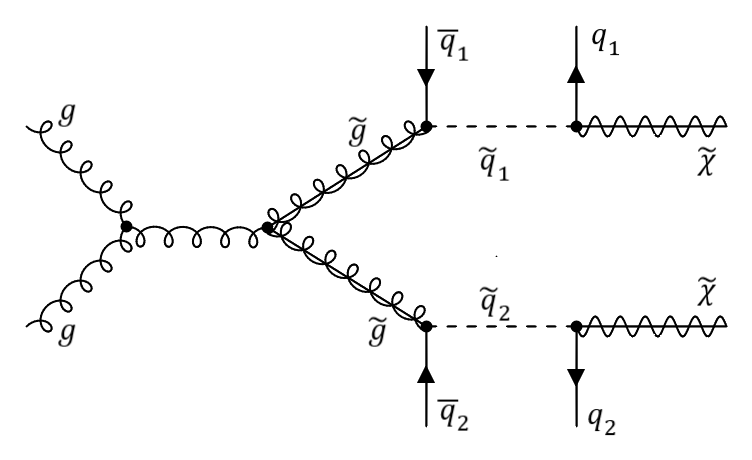}}
\end{center}
\caption{Left panel: Conventional sfermion decay processes, like the one shown here, do not exist in the present scenario 2, because each vertex must involve two sfermion fields and two other bosonic fields. The conventional schemes for detecting sfermions rely on their being produced in collisions through processes that largely do not exist in the present scenario, and then, more importantly, decaying through processes that are entirely disallowed. This implies that new detection schemes are required, and that squarks with masses $\sim$ 1 TeV may exist even though they have not previously been identified. Right panel: Gluino decay processes, like the one shown here, do not exist, because a decay would require a squark-quark vertex to conserve R-parity and color charge, with first-order squark vertices not allowed in the present scenario.}
\label{gluions}
\end{figure}
Many additional topics are beyond the scope of the present paper, including the interactions of the new $\varphi$ fields and particles with fermions.  Our dark matter higgson was introduced in a non-supersymmetric context, with no interactions except those with gauge bosons, and the interactions of squarks considered above involve only gauge and scalar bosons. There is then gauge invariance, as shown at the end of Section~\ref{scalars} (and is evident in (\ref{eq50})). But if the higgsons and redefined sfermions are themselves to be protected from strong divergences due to radiative corrections, it is necessary that they be coupled to gauginos (and, for the sfermions, higgsinos). An $f-\varphi-f$ coupling, with $f$ representing a single fermion field, would satisfy conservation of all charges and invariance under infinitesimal gauge transformations, but not finite gauge transformations. To have full gauge invariance, it appears that $\varphi$ must be a 2-component scalar boson field in this context. 
%(The components must be joined in (\ref{e8y}) to constitute a proper scalar amplitude mode excitation of the vacuum field $\overline{\Phi} $, whose form is itself dictated by the requirement that the vacuum be Lorentz invariant.) 
Then full gauge invariance requires that interactions with fermions have the form $f_a \varphi_a f_a' ,  f_b \varphi_b f_b' $, where $\varphi_a$ and $\varphi_b$ are the two components of $\varphi$, which still transforms as a scalar field under coordinate and Lorentz transformations. (The ``glue'' that holds $\varphi_a$ and $\varphi_b$ together is ultimately the requirement that the vacuum and its excitations be Lorentz invariant, so that the vacuum fields $\overline{\Phi} ^{r} $ and their amplitude modes containing 
$\varphi_a^r$, $\varphi_b^r$ must have the forms (\ref{same}) and (\ref{e8y}).) 
For example, in this extended version of scenario 2, a sufficiently massive redefined squark can decay through this coupling into two gluinos and a quark-antiquark pair, but a single gluino still cannot decay. The basic picture is then changed somewhat, because the new $\varphi$ scalar boson fields must be treated as 2-component objects in these exotic processes. The phenomenology in \cite{susy2025b} is extended with the inclusion of these additional processes, but is still valid for the superpartner masses that were assumed (with, e.g.,  the mass of the lightest squark assumed to be below twice the mass of the lightest gaugino). An $\approx 70$ GeV higgson is still stable if the lightest relevant gaugino mass is above $\approx 35$ GeV.

Gluinos will presumably hadronize into color-neutral complexes, which can again be a minor part of the dark matter. It appears that such completely stable and colorless R-hadrons are consistent with experimental and observational limits (whereas various other varieties of R-hadrons have been ruled out by collider and astronomical constraints)~\cite{pdg}.

The absence of decay processes for squarks and gluinos means that the most heavily emphasized processes for detecting conventional SUSY do not exist, and the experimental signatures for these particles are dramatically changed, with a detailed discussion in \cite{susy2025b}. 

\section{\label{conclusion}Conclusion}
The search for supersymmetry is very well motivated~\cite{Baer-Tata,Dreiner,Nath,Mohapatra,Drees,Binetruy,Djouadi,Kane-susy,Haber-Kane,Arkani-Hamed-Giudice,Ellis-Olive,Baer-Barger-Tata,Roszkowski-2018,Baer-Barger-2020,Tata-2020,3,pdg}, and it has been a major mystery that not a single one of all the many superpartners has yet been discovered. The presentation above suggests this may be because the phenomenology is very different from what has been expected (and incorporated in the analysis and simulation of events): In scenario 2 gluinos are stable particles, which can hadronize but not decay, so they can be detected only as missing energy accompanied by jets or electroweak particles. Squarks also cannot decay through the conventional channels, and will also exhibit an unconventional phenomenology in this scenario.

Calculations of cross-sections and observables have been performed for collider detection of both the dark matter particle and various superpartners, and the results have been or will be published separately~\cite{DM2025,susy2025b}. In ~\cite{DM2025} it was found that, with optimal cuts, the $\approx 70$ GeV dark matter WIMP should be detectable at the high-luminosity LHC, perhaps after only two years with an integrated luminosity of 500 fb$^{-1}$.
In~\cite{susy2025b}, it was found that (with optimal cuts) the predictions for a 1500 GeV squark or 1500 GeV gluino can also be tested at the high-luminosity LHC, at slightly above the 5$\sigma$ level, after 12 years or 3000 fb$^{-1}$ of integrated luminosity, with less required for lower masses; and that weakly interacting superpartners with masses $\sim 400$ GeV or less can be detected above the $5 \sigma$ level at the planned 100 TeV collider (again with optimal cuts).

\appendix

\section{\label{sec:appA}First steps in transformation from primitive to physical boson fields}

The initial gauge group is assumed to be Spin(10), with spin 1/2 fermions belonging to spinorial $32=16+ \overline{16}$ representations (standard model fermions plus new predicted ones), vectorial $10 = 5 + \overline{5}$ representations (higgsinos), and the adjoint $45$ representation  (gauginos). The fundamental covariant derivative is $D_{\mu } = \partial_{\mu} - i A_{\mu}$,  where the coupling constant is absorbed into the gauge potentials $A_{\mu}=A_{\mu}^i t^i$ and the generators $t^i$ are treated as operators. The field strength tensor is $F_{\mu \nu} $, with $\mu, \nu = 0,1,2,3$. Summations are implied over coordinate and gauge-field indices like $\mu$, $k$, and $i$, but not  labels of other fields like $r$. The metric tensor has the form $diag \left(-1, 1, 1, 1 \right)$. In $\sigma^{\mu}$ and  $\overline{\sigma }^{\mu}$, the $\sigma^k$ are Pauli matrices, $\sigma^0$ is the $2 \times 2$ identity matrix, $\overline{\sigma }^{0}=\sigma ^{0}$, and $\overline{\sigma }^{k}=-\sigma ^{k}$. 

All spin 1/2 fields are taken to be initially left-handed. But we will now show that a given left-handed field $\psi _L$ can be transformed into a right-handed charge-conjugate field $\psi_R$, with a result that is standard for fermions but with a minus sign acquired for bosons:
\begin{align}
 i \psi_L^{\dag } \, \overline{\sigma} ^{\mu} D_{\mu } \, \psi _L
\;  \longrightarrow \;  \pm \,  i \psi_R^{c \, \dag} \, \sigma ^{\mu} D_{\mu } \, \psi_R^c  \quad , 
\quad  \psi_R^c = - \sigma^2 C \psi_L^{*} \quad , 
\quad  \psi_L = \sigma^2 C \psi_R^{c \, *}
\label{conj}
\end{align}
where $C=C^{\dag}=C^{-1}$ represents charge conjugation and is here treated as an operator. We will use
\begin{align}
 \quad \sigma^2 \sigma ^k \sigma^2 = - \sigma ^{k \, *} \quad  \text{or} \quad \sigma^2 \overline{\sigma} ^{\mu} \sigma^2 = \sigma ^{\mu \, *}  \label{A50}
\end{align}
and 
\begin{align}
\quad  C A_{\mu} C = - A_{\mu}^* \; .
\end{align}
%(as in e.g. the treatment on pp. 56-56 of \cite{Nanopoulos-1979} or pp. 36-37 of \cite{Ozer} in the case of a spinorial representation)
In the most fundamental convention, the potentials $A_{\mu}^i$ are real and the generators $ t^i $ Hermitian, with 
\begin{align}
C A_{\mu}^i  C = A_{\mu}^i \quad , \quad C  t^i C = - t^{i \, *} \; .
\end{align}
(In the present formulation for path integral quantization using classical fields, $A_{\mu}^i $ is a real number; in canonical quantization it becomes a Hermitian operator, containing the destruction and creation operators for gauge bosons of species $i$.)

The action in (\ref{eq701}) for a single $r$ is equivalent to (with an implied summation over $\alpha$)
\begin{align}
2 {\cal L}^L  &=  i \psi _{L}^{\dagger } \, \overline{\sigma} ^{\mu} D_{\mu } \,  \psi _L   + h.c. \label{eqA10} \\
              &= i \left( \sigma^2 C \psi_R^{c \, *} \right)^{\dag} \, \overline{\sigma} ^{\mu} D_{\mu } \,  \left( \sigma^2 C \psi_R^{c \, *} \right)  + h.c. \\
             &=  i \, \psi_R^{c \, T} \, C \sigma^2 \overline{\sigma} ^{\mu} \sigma^2 D_{\mu } \, C \psi_R^{c \, *}  + h.c. \\
               &= i \, \psi_{R \, \alpha}^c \, \left( \sigma ^{\mu \, *} C D_{\mu } C \,  \psi_R^{c \, *} \right)_{\alpha} + h.c. \\
             &= \mp \, i \left( \sigma ^{\mu \, *} C D_{\mu } C \,  \psi_R^{c \, *} \right)_{\alpha} \psi_{R \, \alpha}^c  + h.c. 
                                                      \quad \text{upper\ sign\ for\ anticommuting\ fermion\ fields}  \\
              &= \mp  \, i \left( \sigma ^{\mu \, *} \left(  \partial_{\mu} - i C A_{\mu} C \right) \,  \psi_R^{c *} \right)_{\alpha} \psi_{R\, \alpha}^c  + h.c. \\
               &= \mp  \, i \left( \sigma ^{\mu \, *} \left(  \partial_{\mu} + i A_{\mu}^* \right) \,  \psi_R^{c \, *} \right)_{\alpha} \psi_{R\, \alpha}^c + h.c. \\
               &= \mp \, i \left( \sigma ^{\mu } D_{\mu} \,  \psi_R \right)^{c \, *}_{\alpha} \psi_{R\, \alpha}^c  + h.c. \\
               &= \mp  \, i  \left( \sigma ^{\mu } D_{\mu } \,  \psi_R^c \right)^{\dag}  \psi_{R}^c  + h.c. \\
               &= \pm \, i \psi_R^{c \, \dag} \sigma ^{\mu} D_{\mu }  \psi_R^c   + h.c.       \label{eqA20}                
\end{align}
% not needed because conjugate field is handled differently
%\begin{align}
%2 \overline{\cal L}^L  &=  \psi _{L}^{\dagger } \, D^{\mu} D_{\mu } \,  \psi _L   + h.c. \label{eqA10x} \\
%              &= \left( \sigma^2 C \psi_R^{c \, *} \right)^{\dag} \, D^{\mu} D_{\mu } \,  \left( \sigma^2 C \psi_R^{c \, *} \right)  + h.c. \\
%             &= \psi_R^{c \, T} \, C \sigma^2 \sigma^2 D^{\mu} D_{\mu } \, C \psi_R^{c \, *}  + h.c. \\
%               &=\psi_{R \, \alpha}^c \, \left(  C D^{\mu}D_{\mu } C \,  \psi_R^{c \, *} \right)_{\alpha} + h.c. \\
%             &=  \left( C D^{\mu}D_{\mu } C \,  \psi_R^{c \, *} \right)_{\alpha} \psi_{R \, \alpha}^c  + h.c. \\
%              &=  \left(  \left(  \partial_{\mu} - i C A_{\mu} C \right) \left( \partial_{\mu} - i C A_{\mu} C \right) \,  \psi_R^{c \, *} \right)_{\alpha} \psi_{R\, \alpha}^c  + h.c. \\
%               &=   \left(  \left(  \partial^{\mu} + i A^{\mu \, *} \right)\left(  \partial_{\mu} + i A_{\mu}^* \right) \,  \psi_R^{c \, *} \right)_{\alpha} \psi_{R\, \alpha}^c + h.c. \\
%                &=  \left(  \left(  \partial^{\mu} - i A^{\mu } \right)\left(  \partial_{\mu} - i A_{\mu} \right) \,  \psi_R^{c } \right)^{*}_{\alpha} \psi_{R\, \alpha}^c + h.c. \\
%                 &=   \left(  \left(  \partial^{\mu} - i A^{\mu } \right)\left(  \partial_{\mu} - i A_{\mu} \right) \,  \psi_R^{c } \right)^{\dag} \psi_{R}^c + h.c. \\
%               &= \psi_R^{c \, \dag}  D^{\mu} D_{\mu }  \psi_R^c   + h.c.       \label{eqA20x}                
%\end{align}
Within the action $S^L = \int d^{4}x \, {\cal L}^L $, the second term of (\ref{eqA20} ) (represented by $h.c.$) gives the same contribution as the first after an integration by parts (with boundary contributions neglected),  so we can write (\ref{eq700})-(\ref{eq701}) as
\begin{align}
S_{fb} = \sum_r S^r_{f} + \sum_r S^r_{b}
\end{align}
with either 
\begin{align}
S^r_{f} = \int d^{4}x \, i \, \psi _{f \, L}^{r \, \dagger } \, \overline{\sigma}  ^{\mu} D_{\mu } \, \psi _{f \, L}^r  \quad , \quad S^r_{b} = \int d^{4}x \, i \, \psi _{b \, L}^{r \, \dagger } \, \overline{\sigma}  ^{\mu} D_{\mu } \, \psi _{b \, L}^r 
\end{align}
or
\begin{align}
S^r_{f} = \int d^{4}x \, i \, \psi _{f \, R}^{r \, c \, \dagger } \, \sigma  ^{\mu} D_{\mu } \, \psi _{f \, R}^{r  \, c } \quad , \quad S^r_{b} = - \int d^{4}x \, i \, \psi _{b \, R}^{r  \, c \, \dagger } \, \sigma  ^{\mu} D_{\mu } \, \psi _{b \, R}^{r \, c}   \; .\label{minus}
\end{align}
Since fermions are treated just as in standard physics, we now focus exclusively on bosons in an $N + \overline{N}$ representation.
Let us leave the boson fields of the $N$ unchanged and left-handed, but change all those of the  $\overline{N}$ to right-handed via the above procedure. There are then N pairs of these new 2-component fields $\overline{\psi}_{b}^{r}$ with the same gauge quantum numbers, where one is left-handed and unchanged and the other is right-handed with the minus sign of (\ref{minus}):
\begin{align}
S_{b} &= \int d^{4}x \, \overline{\mathcal{L}}_{b} \quad , \quad 
\overline{\mathcal{L}}_{b} = \overline{\psi}_{b}^{\dagger } \left( x \right) \,A \, \overline{\psi}_{b} \left( x \right)
= \sum_{r} \overline{\psi}_{b}^{r \, \dagger } \left( x \right) \, A_{r} \, \overline{\psi}_{b}^{r} \left( x \right) \quad , \quad  r=1,2, ..., 2N  \label{eq702} \\
A_{r} &=   i \, \overline{\sigma} ^{\mu} D_{\mu } \quad \text{for\ field\ from\ N\ representation}  \label{eq703} \\
A_{r} &= -  i \, \sigma ^{\mu} D_{\mu } \quad \text{for\ field\ from\ $\overline{N}$ representation}. \label{eq704} 
\end{align}
$A$, with components $A_{r r'} = \delta_{r r'} A_r $, is Hermitian, so it has a complete orthonormal set of eigenfunctions:
\begin{align}
A \, U_i \left( x \right)  = a_i \, U_i \left( x \right) \; . \label{B2} 
\end{align}
Each multicomponent eigenfunction $U_i$ can be taken to be given by a 2-component spinor $u^{\sigma}_s \left( x \right) $ which has well-defined gauge quantum numbers, 4-momentum $p_{\mu}$ -- or frequency $\omega$ and 3-momentum $\vec{p}$ -- and 
helicity $\sigma=+$ or $-$, with the other components equal to 0:
\begin{align}
u^{\sigma}_s \left( x \right) &=  u^{\sigma}_s \left( 0 \right)  e ^{ i p_{\mu} x^{\mu} } = u^{\sigma}_s \left( 0 \right)  e ^{ - i \omega x^{0} + i \vec{p} \cdot \vec{x} } \\
A_{\mu} u^{\sigma}_s \left( x \right)  &= \underline{a}_{\, \mu}^s u^{\sigma}_s \left( x \right) \quad , \quad s, \sigma   \leftrightarrow i \label{B2b} \; .
\end{align}
If $s$ corresponds to the N representation, this gives
\begin{align}
i \overline{\sigma} ^{\mu} D_{\mu } u^{\sigma}_s \left( x \right)  &=  \left[  \left( \omega +  \underline{a}_{\, 0}^s \right) + \left( \vec{p} -  \underline{\vec{a}}^s \right) \cdot \vec{\sigma} \right] u^{\sigma}_s \left( x \right) \label{B2c}  \\
&=  \left(  \tilde{\omega} \, \pm \, \tilde{p}  \right) u^{\sigma}_s \left( x \right) \label{B2d}  
\end{align}
where the upper (lower) sign holds for the spinor which has positive (negative) helicity, and
\begin{align}
\tilde{\omega} = \omega +  \underline{a}_{\, 0}^s \quad , \quad \tilde{p} = |\vec{p} -  \underline{\vec{a}}^s | \label{B2e} 
\end{align}
%(with $\underline{a}_{\, 0} = - \, \underline{a}^{0} $ according to the present convention for the metric tensor), 
so that 
\begin{align}
a_i = a^{\sigma}_s =  \tilde{\omega} \, \pm \, \tilde{p}
\end{align}
where the upper sign holds for $\sigma=+$ (and the lower sign for $\sigma=-$).

If $r$ corresponds to the $\overline{N}$ representation, with $i \, \overline{\sigma} ^{\mu} \rightarrow -  i \, \sigma ^{\mu} $, the result is instead
\begin{align}
- i \sigma ^{\mu} D_{\mu } u^{\sigma}_s \left( x \right)  &=  \left[  - \left( \omega +  \underline{a}_{\, 0}^s \right) + \left( \vec{p} -  \underline{\vec{a}}^s \right) \cdot \vec{\sigma} \right] u^{\sigma}_s \left( x \right) \label{B2f}  \\
 &=  \left(  - \tilde{\omega} \, \pm \, \tilde{p}  \right) u^{\sigma}_s \left( x \right) \label{B2g}  
\end{align}
so that 
\begin{align}
a_i = a^{\sigma}_s =  - \tilde{\omega} \, \pm \, \tilde{p}
\end{align}
in this case.

For a given pair of 2-component spinors specified above with the same gauge quantum numbers (of the N such pairs), there are then 4 eigenfunctions, as listed in Table 1: the 2-component spinor from the N (16 or 5) can have helicity $+$ or $-$, and the same is true of the spinor from the $\overline{N}$ ($\overline{16}$ or $ \overline{5}$). 

As the first major step in the transformation to physical fields, we wish to transform $S_{b}$ to the form
\begin{align}
S_{b} = \int d^{4}x \, \left( \overline{\mathcal{L}}_{\phi} + \overline{\mathcal{L}}_{F} \right) \quad & , \quad 
\overline{ \mathcal{L}}_{\phi} = \overline{\phi} ^{\dag } \left( x \right) B \, \overline{\phi} \left( x \right)  
 \label{phi-F}\\
B = D^{\mu } D_{\mu }   \label{B33}
\end{align}
where  $\overline{\mathcal{L}}_{F}$ involves products of nondynamical auxiliary fields, as specified below.

$B$ also has a complete orthonormal set of eigenfunctions:
\begin{align}
B \, V_i \left( x \right)  = b_i \, V_i \left( x \right)   \; .   \label{B4}
\end{align}
(Each of the $2N$ components of $U_i$ or $V_i$  is itself a spinor with 2 complex components. For a fixed 4-momentum, $A$ and $B$ then each have 4 eigenfunctions for a given set of gauge quantum numbers, as indicated in Table 1.)
We wish to choose the eigenstates of $B$ to be essentially the same as those of $A$, in the sense that each $V_i$ is given by a 2-component spinor with the same gauge quantum numbers and the same 4-momentum (frequency and 3-momentum) as its progenitor in $U_i$. This means that each of the 4 modes in Table 1 for $A$ with a fixed $\tilde{\omega}$ and $\tilde{p}$ has to be matched to a corresponding mode for $B$ with the same $\tilde{\omega}$ and $\tilde{p}$, although the eigenvalues will, of course, be different for the different operators:
\begin{align}
v^{\sigma}_s\left( x \right) &=  v^{\sigma}_s\left( 0 \right)  e ^{ i p_{\mu} x^{\mu} } = v^{\sigma}_s\left( 0 \right)  e ^{ - i \omega x^{0} + i \vec{p} \cdot \vec{x} } \label{B4a}  \\
B v^{\sigma}_s\left( x \right)  &= b_{s}^{\sigma} v^{\sigma}_s \left( x \right) \quad , \quad s, \sigma   \leftrightarrow i  \label{B4b}  
\end{align}
with the other components of $V_i = V^{\sigma}_s$ equal to 0.
Here $\sigma$ is taken below to label the spin orientation of $v^{\sigma}_s$.
The special case
\begin{align}
b_i = b_{s}^{\sigma}  = \tilde{\omega}^2 - \tilde{p}^2  \quad \mathrm{if} \; A_{\mu} \; \mathrm{is\ constant} \label{B4c}
\end{align}
is emphasized in Table 1, but, as pointed out there, the results demonstrate that the required matching can always be accomplished, so that each of the fields $\phi_{\sigma}  \left( x \right) , F_{\sigma}  \left( x \right)$ defined below can be represented by a complete set of states.

With
\begin{align}
\overline{\psi}_{b} \left( x \right) = \sum_i \overline{\psi}_i \, U_i \left( x \right)
\end{align}
(\ref{eq702}) can be rewritten as
 \begin{align}
\overline{S}_b =  \sum_i \overline{\psi}_i ^{\dag} a_i \overline{\psi}_i \; .
\end{align}
Now define
 \begin{align}
\overline{\phi}_i &= \left( a_i / b_i \right) ^{1/2} \overline{\psi}_i \quad \; \text{if}  \; \; b_i / a_i > 0 \label{B19}\\
\overline{F}_i &=  \left( \left| a_i \right| \right) ^{1/2} \overline{\psi}_i \quad \; \text{if}  \; \; b_i / a_i < 0 \label{B20}
\end{align}
where each $b_i$ is matched to a corresponding $a_i$ in the way specified below, so that
 \begin{align}
S_b =  \sum_i \overline{\phi}_i ^{\dag} \, b_i \, \overline{\phi}_i \ + \sum_i \sgn \left( {a_i} \right) \overline{F}_i^{\dag} \overline{F}_i   \label{B70} 
\end{align}
where the limitations on these summations are defined by (\ref{B19}) and (\ref{B20}). 

For each case in Table 1, one of the two states labeled $\phi$ or $F$ can be rotated to have spin up, and the other to have spin down, with no change in the action. These 2-component spinors can then be taken to be the $v^{\sigma}_s\left( x \right)$ of (\ref{B4a})-(\ref{B4b}), giving the multicomponent eigenfunctions $V_i =  V^{\sigma}_s \left( x \right)$. To avoid confusion, let us write $\sigma = \uparrow, \downarrow$ respectively for the spin up and down states labeled $\phi$ in Table 1, and $\Uparrow, \Downarrow$ for the spin up and down states labeled $F$. Then the general multicomponent dynamical fields can be represented as
 \begin{align}
\phi_{\uparrow} \left( x \right) = \sum_s \overline{\phi} _s ^{\uparrow} V^{\uparrow}_s \left( x \right) \quad , \quad
\phi_{\downarrow} \left( x \right) = \sum_s \overline{\phi} _s ^{\downarrow} V^{\downarrow}_s \left( x \right) \quad , \quad s=1,2, ..., N
\end{align}
with components $\phi ^r_{\uparrow} \left( x \right)$ or $\phi ^r_{\downarrow} \left( x \right)$, $r=1, 2, ..., N$, and $S_b$ can now be written as 
\begin{align}
S_b &= S_{\phi} + S_F \\
S_{\phi} &= \int d^{4}x \left( \phi_{\uparrow} ^{\, \dag } \left( x \right)  B \, \phi_{\uparrow}  \left( x \right)  + \phi_{\downarrow} ^{\, \dag } \left( x \right)  B \, \phi_{\downarrow}  \left( x \right) \right) \label{gen} \\
&= \int d^{4}x \, \sum_{r}  \left( \phi_{\uparrow} ^{r \, \dag } \left( x \right)  B \, \phi_{\uparrow}^r   \left( x \right)  + \phi_ {\downarrow} ^{r \, \dag } \left( x \right)  B \phi_{\downarrow}^r \left( x \right) \right) \; .
 \end{align}

$S_F$ is the nondynamical action of the auxiliary fields with terms $\overline{F}_s ^{\Uparrow} V^{\Uparrow}_s \left( x \right)$ and $\overline{F}_s ^{\Downarrow} V^{\Downarrow}_s \left( x \right)$. They will not be considered further because they are not relevant to the main results obtained here.
\begin{center}
\begin{tabular}{ |p{1.9 cm}| |p{1.6cm}| p{0.9cm}| p{0.8cm}| |p{2.3cm}|  p{1.2cm}|  p{1.2cm}|  p{1.5cm}|  }
 \hline
  \multicolumn{8}{|l|}{\hspace{2.2cm}modes of A reassigned to modes of B } \\
 \hline
 $\; \tilde{\omega}$ and   $\tilde{p}$ & \; \, rep & L, R &\hspace{0.015cm}  hel & $A$ eigenvalue & $A$ sign & $B$ sign & $B$ mode \\
 \hline
 $\tilde{\omega} > 0$ & 16 or 5 & L & \;  \,-- & $\tilde{\omega}$ -  $\tilde{p}$  & +  & + & $\phi $ \\
$|\tilde{\omega} | >  \tilde{p} $ & 16 or 5 & L & \,   +   & $\tilde{\omega}$ +  $\tilde{p}$    & + & +  & $\phi $ \\
& $\overline{16}$ or $\overline{5}$ & R & \,  + &  - $\tilde{\omega}$ +  $\tilde{p}$   & \,--  & + & \footnotesize{F} \\
& $\overline{16}$ or $\overline{5}$ & R & \;   \,--  & - $\tilde{\omega}$ -  $\tilde{p}$  & \,-- & + & \footnotesize{F} \\
\hline
$\tilde{\omega} > 0$ & 16 or 5  & L & \;  \,--  & $\tilde{\omega}$  -   $\tilde{p}$  &  \,--  &  \,--  & $\phi $ \\
 $|\tilde{\omega} | <  \tilde{p} $ &16 or 5 & L & \, +  & $\tilde{\omega}$ +  $\tilde{p}$   & + &  \,--  & \footnotesize{F} \\
& $\overline{16}$ or $\overline{5 }$  & R & \,  +  & - $\tilde{\omega}$+  $\tilde{p}$  & + &  \,--  & \footnotesize{F} \\
& $\overline{16}$ or $\overline{5 }$ & R & \;   \,--   &  - $\tilde{\omega}$ -  $\tilde{p}$   &  \,--  &  \,--  & $\phi $ \\
\hline
$\tilde{\omega} < 0$ & 16 or 5   & L & \;  \,--   & $\tilde{\omega}$ -   $\tilde{p}$  &  \,--   & + & \footnotesize{F}\\
$|\tilde{\omega} | >  \tilde{p} $ & 16 or 5 & L &  \,  +  & $\tilde{\omega}$ +  $\tilde{p}$    &  \,--  & + & \footnotesize{F}  \\
& $\overline{16}$ or $\overline{5 }$ & R & \,  +  & - $\tilde{\omega}$ +  $\tilde{p}$  & +  & + & $\phi $ \\
& $\overline{16}$ or $\overline{5 }$ & R& \;   \,-- &  - $\tilde{\omega}$ -  $\tilde{p}$   & + & + & $\phi $ \\
\hline
$\tilde{\omega} < 0$ & 16 or 5  & L & \;  \,--   & $\tilde{\omega}$ -  $\tilde{p}$  &  \,-- &  \,-- & $\phi $ \\
$|\tilde{\omega} | <  \tilde{p} $ & 16 or 5 & L & \,  +  & $\tilde{\omega}$ +  $\tilde{p}$   & + &  \,--  & \footnotesize{F} \\
& $\overline{16}$ or $\overline{5 }$ & R & \,  +  & - $\tilde{\omega}$ +  $\tilde{p}$ & + &  \,--  & \footnotesize{F} \\
& $\overline{16}$ or $\overline{5 }$ & R & \;   \,--   &  - $\tilde{\omega}$ -  $\tilde{p}$   &  \,--  &  \,-- & $\phi $ \\
 \hline
\end{tabular} \\
\end{center}
\noindent
Table 1. Here we consider how the eigenstates $u_i$ and $v_i$ of $A$ and $B$ -- the operators defined in (\ref{eq703})-(\ref{eq704}) and (\ref{B33}) -- can be matched. In the case of $A$, any one of the 16 or 5 pairs in the full $32 \times 32$ or $10 \times 10$ vector $U_i$ of (\ref{B2}) consists of two fields with the same gauge quantum numbers: a left-handed field from the 16 or 5 representation and a right-handed field from the $\overline{16}$ or $\overline{5 }$. For a specific $\omega$ 
and $\vec{p}$ in (\ref{B2e}), there are two eigenstates for the left-handed field and two for the right-handed field, corresponding to positive and negative helicities, labeled $+$ and $-$ respectively. These eigenvalues and their signs are given in the columns labeled ``$A$ eigenvalue'' and ``$A$ sign'', for the four possible combinations of $\tilde{\omega}$ and $\tilde{p} $ which yield different signs. As noted in (\ref{B4c}), in the special case that $A_{\mu}$ is constant, the corresponding eigenvalues of $B$ are $ \tilde{\omega}^2 - \tilde{p}^2 =\left( \left| \tilde{\omega}\right| +\tilde{p} \right) \left( \left| \tilde{\omega} \right| - \tilde{p} \right)$, with the sign of $b_i $ always determined by $\left( \left| \tilde{\omega} \right| - \tilde{p} \right)$, as indicated in the column ``$B$ sign''. If the signs for $A$ and $B$ agree, the matched eigenstate of $B$ is labeled $\phi$. If they do not agree, the matched state is labeled $F$. It is remarkable that in every case two matches are obtained for $\phi$ and two for $F$. Moreover, this matching holds regardless of the sign for $B$, since if this sign is reversed, for any of the 4 cases shown in the table, there are still two matching eigenvalues for $\phi$ and two for $F$  (since $A$ always has two $+$ and two $-$ eigenvalues). This implies that each of the fields $\phi_{\uparrow} \left( x \right)$, $F_{\uparrow} \left( x \right)$, $\phi_{\downarrow} \left( x \right)$, $F_{\downarrow} \left( x \right)$ of (\ref{gen}) can always be represented by a complete set of states.

\bigskip\noindent
\textbf{Acknowledgements}

\medskip\noindent
The Feynman diagrams of Figs.~1 and 2 were prepared by Bailey Tallman.


\begin{thebibliography}{999}
\bibitem{quote1}The central principle and motivations in the present work are the same as defined in the preface of \cite{Dreiner}: ``Supersymmetry is the principle that fermions and bosons are related by a symmetry\,.... From a practical point of view, supersymmetry provides a way of tremendously reducing the hierarchy problem associated with the Higgs field, and nontrivially it also provides avenues of approach to the other new outstanding physics puzzles noted above, especially the dark matter and unification ideas.''
\end{thebibliography}
\end{document}